\documentclass[preprint2,twoside]{hwo}

\usepackage{xcolor}
\usepackage{natbib}
\usepackage{array}
\newcolumntype{x}{>{\raggedright\arraybackslash}p{0.17\linewidth}}

\input{hwo.h}

\begin{document}

\title{\textbf{\LARGE Envisioning the Distance Ladder in the Era of the Habitable Worlds Observatory}}
\author {\textbf{\large Gagandeep Anand,$^{1}$ Meredith Durbin,$^2$ Rachael Beaton$^1$ }}
\affil{$^1$\small\it Space Telescope Science Institute, Baltimore, Maryland, USA}
\affil{$^2$\small\it University of California, Berkeley, CA, USA}

\author{\small{\bf Contributing Authors:} Joseph Jensen (Utah Valley University), Adam Riess (Johns Hopkins University)}

\author{\footnotesize{\bf Endorsed by:}
Howard Bond (Penn State University), Michael Davis (Southwest Research Institute), Chris Impey (University of Arizona), Pierre Kervella (Paris Observatory \& CNRS IRL FCLA), Eunjeong Lee (EisKosmos (CROASAEN), Inc.), Karen Masters (Haverford College), Kristen McQuinn (Space Telescope Science Institute \& Rutgers University), Ilaria Musella (National Institute for Astrophysics (INAF) - Astronomical Observatory of Capodimonte (Naples)), Donatas Narbutis (Institute of Theoretical Physics and Astronomy, Faculty of Physics, Vilnius University, Vilnius, Lithuania),  Chow-Choong Ngeow
(Graduate Institute of Astronomy, National Central University), Javier Pascual-Granado (Institute of Astrophysics of Andalusia (CSIC)), Nicolás Rodríguez-Segovia (UNSW Canberra), Frank Soboczenski (University of York \& King's College London)}

% This section is for ADS Processing.  There must be one line per author. Leave them commented out for the present. They will be included later.
%\paperauthor{Sample~Author1}{Author1Email@email.edu}{ORCID_Or_Blank}{Author1 Institution}{Author1 Department}{City}{State/Province}{Postal Code}{Country}
%\paperauthor{Sample~Author2}{Author2Email@email.edu}{ORCID_Or_Blank}{Author2 Institution}{Author2 Department}{City}{State/Province}{Postal Code}{Country}
%\paperauthor{Sample~Author3}{Author3Email@email.edu}{ORCID_Or_Blank}{Author3 Institution}{Author3 Department}{City}{State/Province}{Postal Code}{Country}

% Please provide entries for the Author index
%\aindex{Author, F.}
%\aindex{Author, S.}
%\aindex{Author, T.}

\begin{abstract}
  The current state-of-the-art cosmic distance ladder requires three rungs---geometric distances, primary indicators, and Type Ia Supernovae---to achieve a 1\% measurement of the Hubble constant $H_0$.
  The Habitable Worlds Observatory will have the sensitivity and resolution to reduce this to a two-step measurement, eliminating the third rung entirely and reaching into the Hubble flow with stellar distance indicators such as Cepheid variables and the tip of the red giant branch alone.
  We discuss the requirements for a program to measure $H_0$ to 1\% with HWO here, including telescope and instrument design considerations. We also comment on the potential of HWO to measure distances to low-mass dwarf galaxies via their RR Lyrae stars. 
 \\
 \\
\end{abstract}

\vspace{2cm}

\section{Introduction}

Galaxy distances are a fundamental astrophysical quantity. Distances to astronomical objects translate between observed quantities (relative system or angular values) and astrophysical quantities (absolute system and physical dimensions) necessary to compare objects across space and time. As such, knowledge of precise and accurate distances are required to set the foundation for most extragalactic science. Despite great advances in the extragalactic distance scale in modern times, distances within $\sim$100 Mpc ($\sim$300 million light-years) remain one of the most difficult quantities to measure directly. 

In addition to providing the basis for measuring nearly all physical parameters of nearby galaxies, distances allow us to probe the evolution of structure over a broad range of environments (from sparse voids to rich galaxy clusters), and perform highly precise and accurate measurements of the expansion rate of the Universe, allowing us to provide strict constraints on cosmological models. 

The latter point is especially relevant given the current state of the ``Hubble tension", a significant disagreement between the model-predicted expansion rate of the Universe at the present day, versus what is actually measured in the local Universe with premier facilities such as Hubble and JWST \citep{2024hct..book.....D}. While the most recent results from these observatories show little sign of easing this tension, several concerns remain. HWO will allow us to make major breakthroughs in our understanding of the physics driving cosmic expansion and the formation and evolution of the Universe as a whole.

This contribution summarizes a Science Case Development Document (SCDD). This SCDD relates to the following topics in the Astro2020 Decadal Survey \citep{2021pdaa.book.....N}, while also covering topics under the purview of the NASA Astrophysics Program Offices for Cosmic Origins and the Physics of the Cosmos:
\begin{itemize}
    \item What physics drives the cosmic expansion and the large-scale evolution of the universe?
    \item How do the histories of galaxies and their dark matter halos shape their observable properties?
    \item What are the most extreme stars and stellar populations?
\end{itemize}

\section{Science Objective}

\subsection{Measuring distances out to 100 Mpc using primary indicators}

The modern distance ladder is precise and accurate (with individual distances routinely achievable at the $\sim5\%$ level), but still requires great effort and detailed observations. Beyond 100 Mpc, we can use an galaxy’s observed velocity and the Hubble constant as a proxy for distance when needing to determine physical parameters, with minimal uncertainty ($\lesssim$5\%) resulting from peculiar velocities due to galaxy-galaxy interactions (although galaxy distances beyond 100 Mpc are still needed for generating maps of peculiar velocity fields themselves). However, within 100 Mpc, uncertainties from peculiar velocities are significant enough that many science cases will require carefully measured distances via resolved stellar distance indicators.  

With respect to the primary goal of measuring the current expansion rate of the Universe, the Hubble Constant is measured in three steps at present. The first is the geometric foundation, where a physical distance is directly measured from an angular size. The best-known of these is trigonometric parallax \citep[e.g., Gaia;][]{2016A&A...595A...4L}, but orbital modeling of eclipsing binary stars and water megamasers around active galactic nuclei are critical for extragalactic distances as well \citep{2019Natur.567..200P, 2020ApJ...891L...1P}. Next are primary indicators, which relate an observed luminosity to a known intrinsic one by the inverse square law of light, and which can be calibrated directly to geometric anchors. These include pulsating variable stars such as classical Cepheids and RR Lyrae, whose pulsation periods correlate with their luminosities, and color-magnitude diagram (CMD) features such as the tip of the red giant branch (TRGB) and J-region asymptotic giant branch (J-AGB), which occur at more-or-less fixed absolute magnitudes at certain wavelengths. The third and last is Type Ia Supernovae (SNe Ia), extremely energetic transient events with roughly constant peak luminosities that can be detected out to great distances in the Hubble flow.

HWO, on the other hand, will have the sensitivity and resolution to measure Cepheid distances out to 100 Mpc. This is far enough into the Hubble flow that peculiar velocities no longer dominate redshift uncertainties. This in turn will enable us to put together a two-step distance ladder (anchors and Cepheids), \emph{bypassing SNe Ia completely} and thereby greatly reducing the total uncertainty budget by dropping the third-rung (SNe Ia) of the present-day distance ladder. SNe~Ia still have a number of unsolved problems lingering over their use \citep[e.g., incomplete understanding of their progenitors, uneven distribution on the sky due to history of ground-based measurements, ground-based photometry with differing instrumental systematics, etc.;][]{2022ApJ...938..111B,2023ApJ...945...84P}.

\begin{figure*}[htb!]
    \centering
    \includegraphics[width=\linewidth]{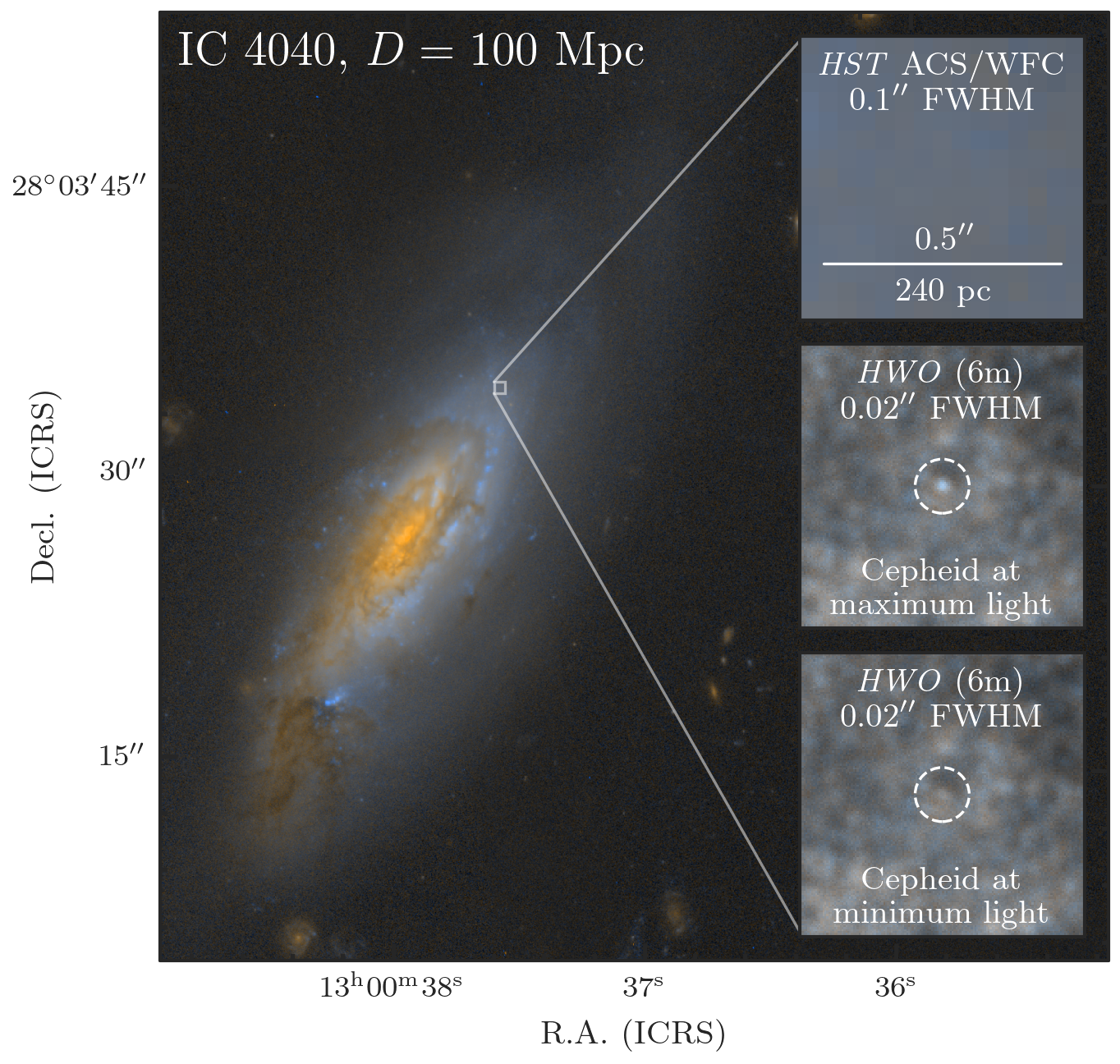}
    \caption{Main panel: false-color HST imaging of IC 4040, a spiral galaxy in the Coma Cluster at 100 Mpc. The upper inset panel zooms in on a small region of the HST image, showing that individual stars cannot be resolved. The lower two panels show a simulation of the same scene imaged with a 6m version of HWO, which resolves the galaxy into stars. The center panel shows a luminous Cepheid at maximum light ($M_V = -6.5$), and the lower the same Cepheid at minimum light ($M_V = -5.1$).}
    \label{fig:cepheid_sim}
\end{figure*}

\subsection{Galaxy parameters \& population-dependent distance techniques}

Classical Cepheids will see the most transformative gains for establishing $H_0$ with HWO. While JWST is transforming the quality of infrared measurements of Cepheids, its blue cutoff is too red to \emph{identify} previously unobserved Cepheids, effectively limiting SNe~Ia host sample sizes to what is reachable by HST and locking in Cepheid samples in HST-surveyed galaxies at their current number. Figure~\ref{fig:cepheid_sim} shows existing HST and simulated HWO observations of a spiral galaxy, IC\,4040, at 100 Mpc in the Coma Cluster. The insets compare HST ACS/WFC to HWO --- where HST cannot resolve a single star, HWO is able to measure the star at both peak and minimum light. Deep, precise photometry across all phases is required to characterize the Cepheid.

Classical Cepheids are pulsating variable stars of intermediate mass that evolve through the classical instability strip while taking its ``blue loop". The classical instability strip marks a set of physical conditions in stellar interiors where the star’s energy generation is unstable to pulsation. The pulsation period is related to several physical parameters of the star during this phase and the empirically defined Leavitt Law relates the period to the mean luminosity of the star \citep{2015pust.book.....C}. With 30 years of observations, starting with the initiation of the HST Key Project, \emph{Cepheids are the de-facto distance indicator to calibrate SNe Ia and anchor cosmological measurements} \citep{2022ApJ...934L...7R}. 

However, Cepheid-based distances are not without significant complications. Cepheids are young stars ($\sim$20-200 Myr) and are only present in galaxies with recent star formation \citep[typically of spiral morphologies;][]{2015pust.book.....C}. Furthermore, Cepheids are relatively vulnerable to systematic effects from crowding and extinction, as young stellar populations tend to be highly clustered and embedded in gas and dust. Lastly, the Leavitt Law also has a metallicity dependence. Within the Milky Way, the metallicity dependence remains challenging to constrain star-to-star due to complications with self-consistent modeling of non-static stellar atmospheres in the already challenging low-density atmospheric regime \citep[e.g.,][]{2021MNRAS.508.4047R}. With current facilities, it is impossible to measure star-by-star metallicities for Cepheids and proxies are used. In the last decade, the investment in other distance indicators have seen a revival as a means to cross-check the Cepheids in light of the Hubble Tension.

\begin{figure*}[htb!]
    \centering
    \includegraphics[width=0.625\linewidth]{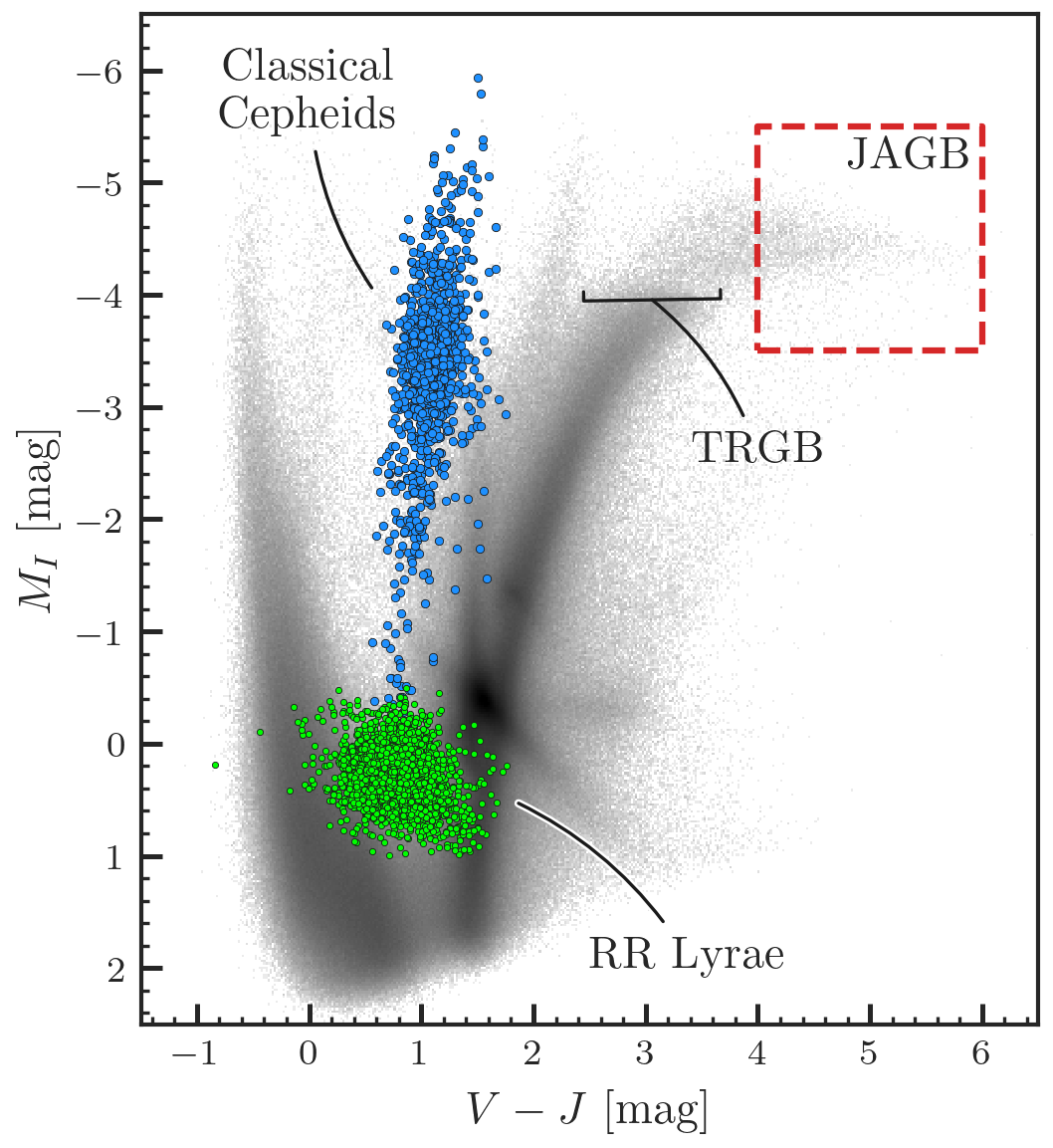}
    \caption{Absolute color-magnitude diagram (CMD) of the Large Magellanic Cloud highlighting the four stellar distance indicators considered here. Data from OGLE \citep{2008AcA....58...89U,2015AcA....65....1U, 2016AcA....66..131S, 2017AcA....67..103S, 2021ApJS..252...23S} and VMC \citep{2011A&A...527A.116C}.}
    \label{fig:cmd}
\end{figure*}

\emph{Several alternative distance indicators with different population characteristics---and therefore independent astrophysical systematics---must be established concurrently} to achieve a self-consistent distance scale applicable to all galaxy morphologies. We discuss three such alternatives here that use resolved stellar populations: the tip of the red giant branch (TRGB), the J-region asymptotic giant branch (J-AGB), and RR Lyrae variables (Figure~\ref{fig:cmd}). 

\subsection{The TRGB and J-AGB for Ancient and Intermediate-aged Populations}

The tip of the red giant branch (TRGB) is the terminal magnitude of the RGB sequence—as low mass stars ($\lesssim$2 $M_{\odot}$) evolve up the RGB, they approach a point where the degenerate helium core reaches a sufficient mass to initiate helium fusion in a runaway process known as the helium flash. The key point here is that the core mass at which this occurs is constant \citep[to within 0.02 $M_{\odot}$;][]{2017A&A...606A..33S}, resulting in a uniform bolometric luminosity of this feature, and hence its standard candle nature. In practice, effects of line blanketing that vary with metallicity result in the need for corrections that vary as a function of wavelength— these corrections (and hence uncertainties) are minimized at $\sim$900 nm, and hence the TRGB has been typically measured in the $I$-band \citep{2018ApJ...866..145H,2018ApJ...861..104H,2018ApJ...852...60J,2023ApJ...954...87W}.

\begin{figure*}[htb!]
    \centering
    \includegraphics[width=0.49\linewidth]{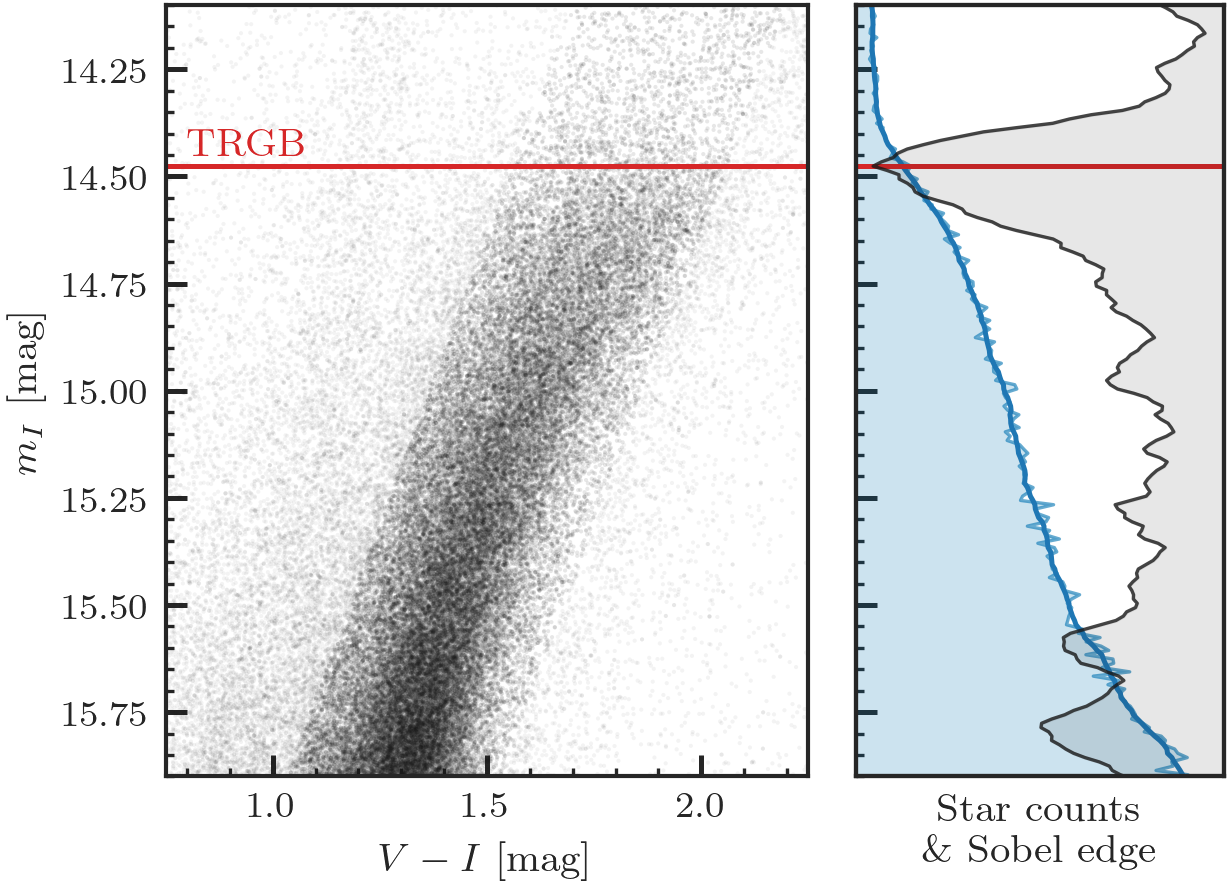}
    \includegraphics[width=0.49\linewidth]{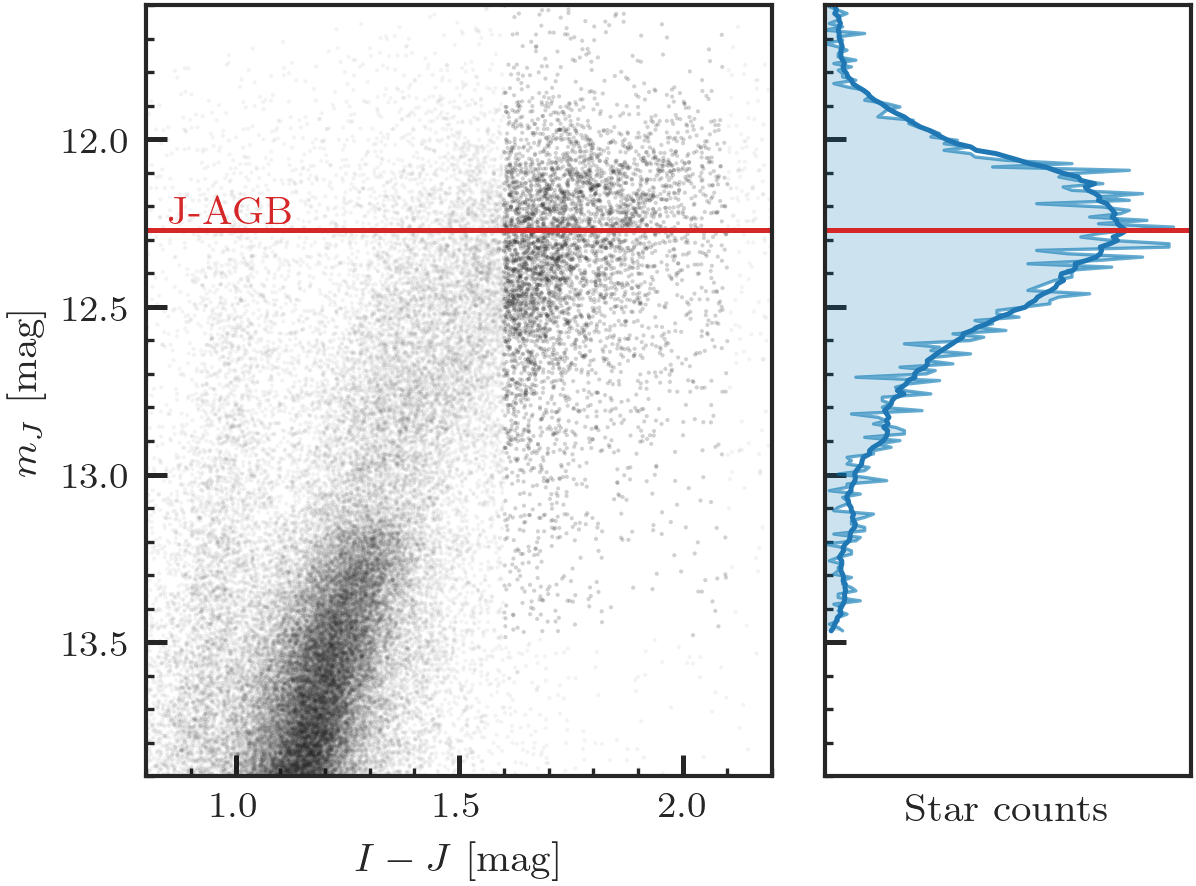}
    \caption{Demonstrations of detecting the TRGB (left) and J-AGB (right) in the LMC via their respective luminosity functions. Data from OGLE \citep{2008AcA....58...89U,2015AcA....65....1U, 2016AcA....66..131S, 2017AcA....67..103S, 2021ApJS..252...23S} and VMC \citep{2011A&A...527A.116C}.}
    \label{fig:trgb_jagb}
\end{figure*}

The TRGB is measured in a color-magnitude diagram by either using an edge-detection algorithm (Figure~\ref{fig:trgb_jagb}, left) along the giant branch luminosity function (LF), or by fully modelling the LF itself. The TRGB is best measured in the halo of a galaxy (e.g., outside of its central disk and in its low density environments), where the impacts of metallicity variations, contamination from younger populations, dust extinction, and photometric crowding are all at a minimum \citep{2016ApJ...832..210B,2019ApJ...885..141B,2021ApJ...906..125J,2022ApJ...932...15A}. Given the differences between these stars and Cepheids, the TRGB serves as an excellent complementary way to measure distances, with distinct astrophysical systematics. Additionally, old ($\gtrsim$5~Gyr) RGB stars are ubiquitous in all manner of galaxies, meaning the TRGB can be used to measure distances to effectively any galaxy with sufficient stellar mass to fully populate the TRGB feature \citep[$>5 \times 10^{6} M_{\odot}$;][]{2019ApJ...880...63M}.

The J-AGB method \citep{2020ApJ...899...67F} uses the mean, median, or mode luminosity of a plume of carbon-rich AGB stars (300 Myr $<$ age $<$ 1 Gyr) as a standard candle (Figure~\ref{fig:trgb_jagb}, right). The method is relatively new, and issues with regards to metallicity, age, and contamination from other stellar populations are still currently being developed in the literature \citep{2024ApJ...966...20L}. Still, initial results from JWST \citep[including but not limited to:][]{2024ApJ...961..132L,2025ApJ...985..182L,2025arXiv250205259L} support the methodology and it has been shown that there is general agreement with Cepheids and TRGB stars \citep{2024ApJ...977..120R}. J-AGB stars are expected to be most plentiful in the outskirts of disks, where they will also be reasonably uncrowded \citep{2022ApJ...933..201L}. Again, the differing systematics provide yet another independent point of comparison for the second rung of the distance ladder.

Early work with JWST \citep{2024ApJ...975..111H,2024ApJ...966...89A} has demonstrated the ability to measure Cepheid, TRGB, and J-AGB distances to galaxies within a singular pointing, which, if fully realized, would be a significant boon for the distance scale. Any inconsistencies in the distance measurements to a single galaxy dervied from these independent tracers would provide evidence towards systematic errors in stellar astrophysics and distances, as opposed to concerns about $\Lambda$CDM, being a source of the Hubble Tension \citep[see e.g.,][]{2024hct..book.....D}. While some early JWST results show minimal reason for a concern among the three distance indicators \citep{2024ApJ...977..120R}, others show the potential for larger differences \citep{2025ApJ...985..203F}. 

In addition to the above mentioned distance indicators (and RR Lyrae below), there are other primary and secondary distance indicators that HWO will help advance. These include oxygen-rich Mira variables, which are luminous variable AGB stars with their own period-luminosity relations. Mira variables have been used to measure $H_0$ \citep[e.g.,][]{2024ApJ...963...83H}, and provide yet another check as a primary distance indicator.

In terms of secondary distance indicators --- those that are not directly anchored by geometrical means and require a primary distance indicator for calibration --- surface brightness fluctuations \citep[SBF;][]{1988AJ.....96..807T,2003ApJ...583..712J,2021ApJ...911...65B} will be measurable to several hundred Mpc with HWO's extraordinary spatial resolving power. HWO measurements will provide dozens, if not hundreds, of Cepheid and TRGB measurements out to 100 Mpc in the same galaxies as SBF, reducing systematic uncertainties in the SBF calibration and establishing the HWO distance ladder far beyond the limit of methods requiring resolution of individual stars. HWO may allow SBF measurements in low surface brightness (LSB) and other dwarf galaxies where redshifts are unavailable, or in cases where somewhat larger uncertainties are acceptable (in exchange for shorter observing times).

\subsection{The RR Lyrae ladder to low-mass galaxies}

RR Lyrae variables are ancient ($>$10 Gyr), low-mass (0.5-0.8 $M_{\odot}$) horizontal branch pulsators, which show period-luminosity relations like Cepheids in the near-IR, and period-Wesenheit (reddening-free magnitude) relations in the optical. Although they are significantly fainter (by 4+ mag) than the other distance indicators we consider here, they are the best---indeed, often the only---available distance indicators for galaxies fainter than $M_V \approx -8$, as such galaxies do not populate the RGB sufficiently to measure a robust TRGB \citep[see e.g.,][]{1995AJ....109.1645M,2019ARA&A..57..375S}, and most are too ancient to host Cepheids or J-AGB stars at all (both of which are more massive stars for this age and are less populated than the RGB). RR Lyrae distances may not play into $H_0$ directly, but they will be critically important for detailed studies of the faint, isolated field dwarfs that LSST is expected to detect throughout the Local Volume \citep{2021ApJ...918...88M}, which are thought to represent the most pristine fossils of the reionization era possible. RR Lyrae will additionally be useful for studying dwarf galaxy populations around massive hosts other than the MW and M31, as well as for analyzing substructure in the halos of massive galaxies, as RR Lyrae are excellent tracers of ancient stellar populations \citep{2009Ap&SS.320..261C,2014ApJ...784...76Y,2022Univ....8..191M,2022ApJ...938..101S}.

\section{Physical Parameters} \label{se}
In this section we discuss steps to produce a distance ladder with HWO observations, with a focus on the terms in the uncertainty budget for $H_0$. 

\subsection{Defining a Cepheid host and anchor sample}

The central question for our experiment is to determine: \emph{How many Cepheid hosts at $\sim$100 Mpc do we need for a 1\% measurement of $H_0$?} This question requires establishing a detailed error budget for the steps in the ultimate $H_0$ measurement. For simplicity, we follow the SH0ES error budget \citep[e.g.,][]{2022ApJ...934L...7R}. We adopt the values for the anchors and that places contraints to determine the number of galaxies needed for a 1\% measurement of $H_0$ with a two-rung HWO distance ladder. 

\subsection{First rung (anchors)} \label{ssec:anchors}

\begin{figure*}[htb!]
    \centering
    \includegraphics[width=\linewidth]{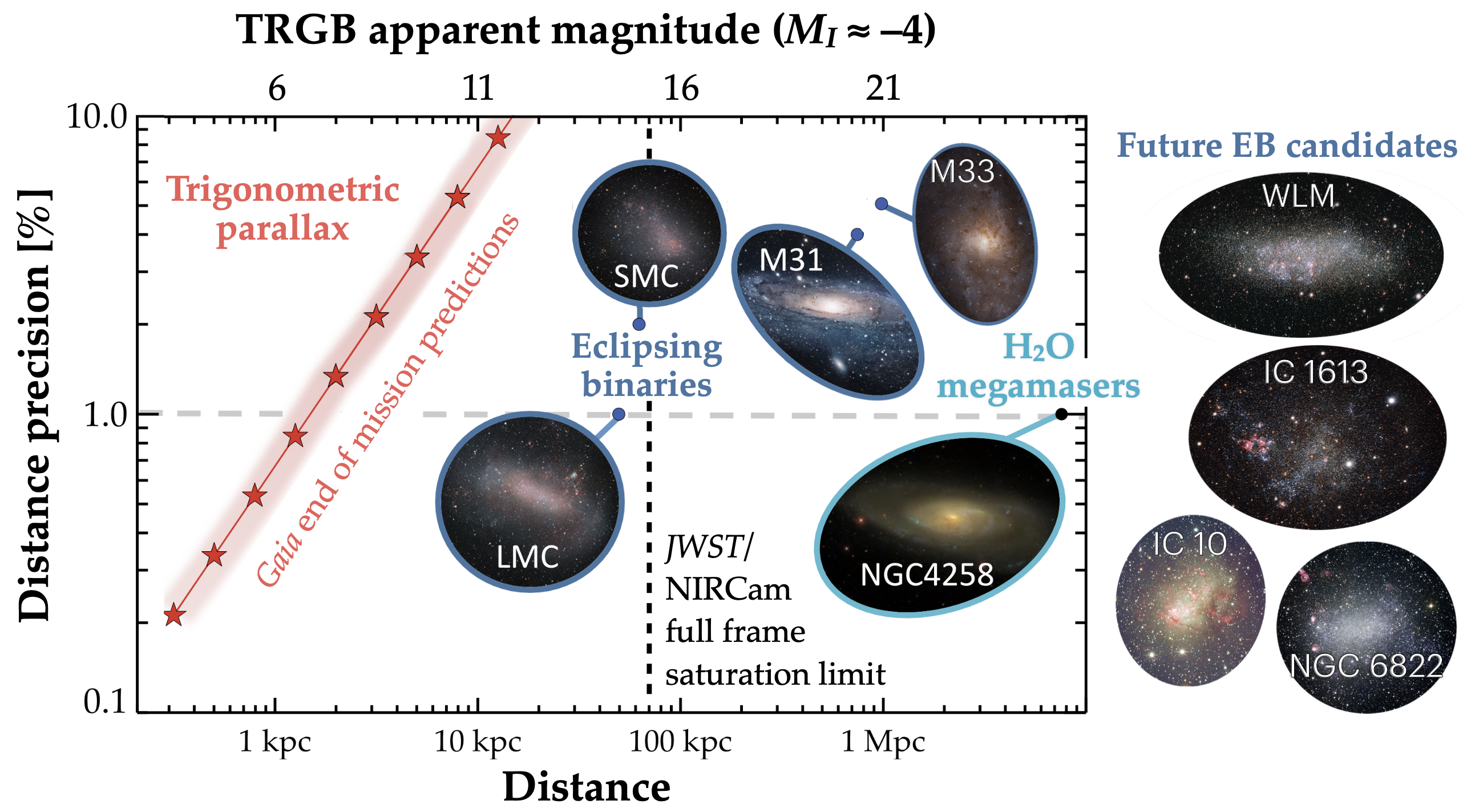}
    \caption{A comparison of state-of-the-art geometric distances that serve to anchor the distance scale and their current relative uncertainties. Additional candidate ($\lesssim$1 Mpc) eclipsing binary hosts to target in future are also pictured. Current EB distances to M31 and M33 rely on early-type stars and are not precise enough to be included in $H_0$ determination \citep{2010A&A...509A..70V,2006ApJ...652..313B}, but we expect this will improve with more sensitive moderate- to high-resolution spectroscopy capable of characterizing late-type EB systems.}
    \label{fig:anchors}
\end{figure*}

The first rung of the HST SH0ES distance ladder has an uncertainty of 0.6\%, with four underlying high-precision geometric anchors (the LMC, SMC, Milky Way, and NGC 4258; see Figure~\ref{fig:anchors}). Assuming the same level of uncertainty could be achieved as the current state-of-the-art in the LMC, and adding at least two additional anchors from late-type detached eclipsing binary systems in the Local Group (M31, M33, and potentially several other Local Group galaxies as well) with the advent of the ELT era \citep[see e.g.,][]{2019BAAS...51c.456B}, we will be able to reduce this uncertainty contribution to 0.5\% or less in the $H_0$ error budget. A caveat is that HWO will need ways to observe nearby Cepheids in the LMC, SMC, and Milky Way, which are quite bright (see Section~\ref{sec:observations} for more details).
% DEB: 2018ApJ...860....1G

\subsection{Second rung}

The uncertainty in the second rung of the proposed two-step distance ladder consists of three subterms:
\begin{itemize}
    \item The uncertainty on the mean of the P-L relation in the Cepheid host galaxies is given by $\sigma_{PL}$/$\sqrt{N}$ = $\sim$2.6\%/$\sqrt{N}$. Each Cepheid host should contain at least 25 Cepheids for sufficient sampling. % Unlike with the SH0ES program
    There is no requirement for these galaxies to contain supernovae, as these galaxy distances can be used to determine $H_0$.
    \item The uncertainty in the P-L slope between the anchor and Cepheid host galaxies, $\sim$0.3\% (conservatively set equivalent to the HST SH0ES value, though this may decrease with the introduction of M31 as an anchor).
    \item A final term is required to account for the residual from peculiar velocity corrections. Even at 80—100 Mpc distances, the peculiar velocity contributions are not negligible and need to be corrected for with peculiar velocity maps \citep[e.g.,][]{2023A&A...670L..15C,2023MNRAS.524.1885W, 2024MNRAS.531...84B}. This term will be of order 2.5\% (150/6,000 km/s) per galaxy. Overall, this term will scale as 2.5\%/$\sqrt{N}$.
\end{itemize}

\subsection{Miscellaneous systematics}

These are additional systematics not contained in the earlier described terms. 
\begin{itemize}
    \item Analysis systematics: These include uncertainties in the Cepheid metallicity correction, as well as uncertainties in the crowding correction procedures with artificial stars. We adopt 0.3\%; this is equivalent to the the term used in the HST-only SH0ES estimates.
    \item A key caveat is that there is likely a systematic uncertainty from local large-scale structure. Results from DESI may help quantify this term in the near future \cite[e.g.,][]{2025arXiv250314745D}. Developing multiple Cepheid-host samples, for example in each quadrant of the sky, would allow for tests of this systematic.
\end{itemize}

\begin{figure*}[htb!]
    \centering
    \includegraphics[width=\linewidth]{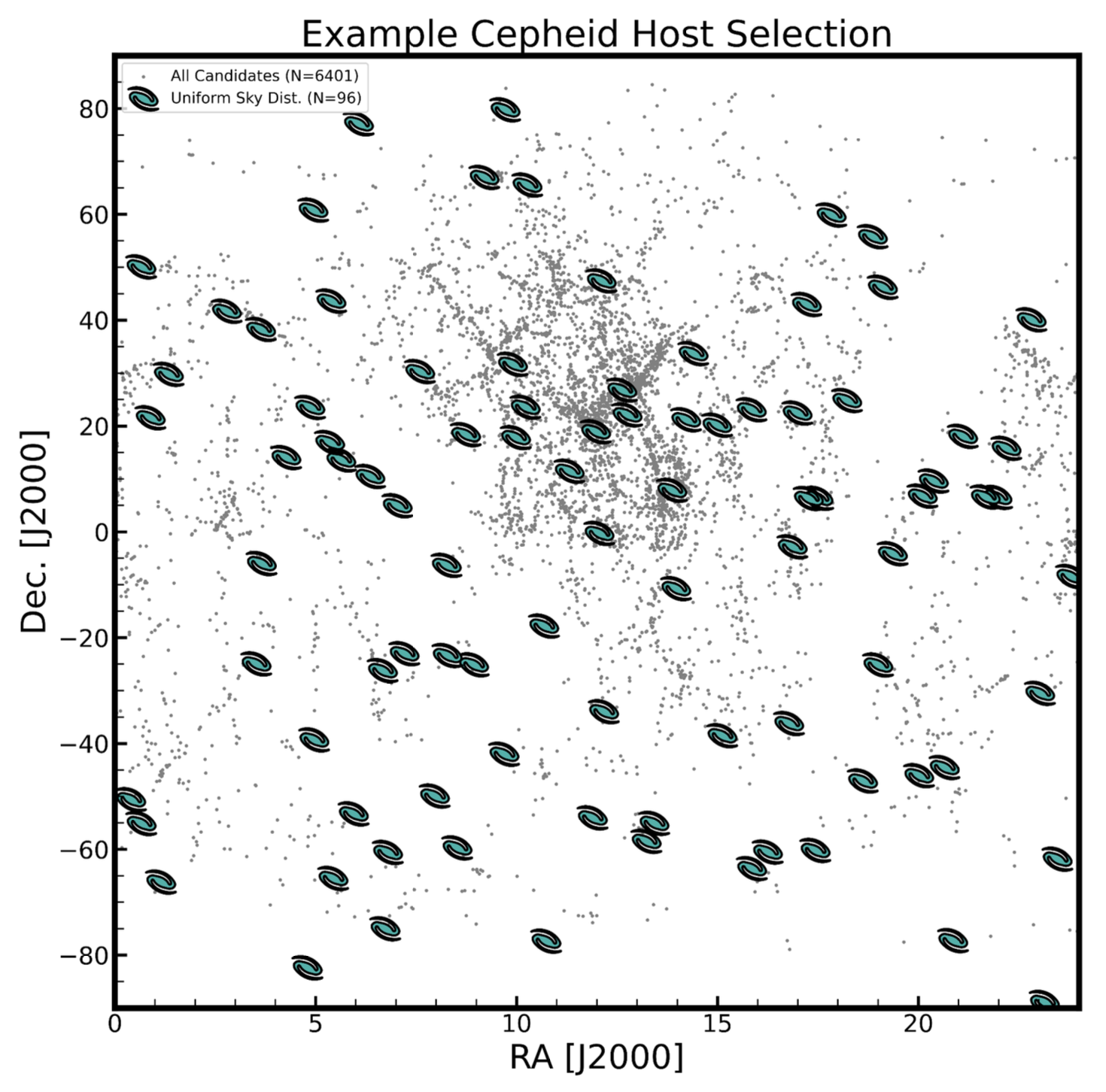}
    \caption{An example Cepheid host galaxy selection for a two-rung, 1\% determination of the Hubble constant in each quadrant of the sky. Candidates were drawn from the HYPERLEDA (\url{https://atlas.obs-hp.fr/hyperleda/}) database of galaxies, with cuts on velocity ($6000 < v < 7500$ km/s), galaxy morphology (Sa through Sd), and inclination ($i<60^{\circ}$).}
    \label{fig:cepheid_sample}
\end{figure*}

\textbf{To achieve a total uncertainty of 1\% in $H_0$, we find that 24 Cepheid host galaxies will be sufficient.} To rigorously test for systematics related to local large-scale structure, the experiment should measure $H_0$ to 1\% in each quadrant of the sky, resulting in $N = 24\times4 = 96$ Cepheid host galaxies, with relatively uniform distribution on the sky (Figure~\ref{fig:cepheid_sample}). 

\begin{table*}[htb!]
    \centering
\begin{tabular}{|x|x|x|x|x|}
\hline
\textbf{Physical Parameter} & \textbf{State of the Art}  & \textbf{Incremental Progress (Enhancing)} & \textbf{Substantial Progress (Enabling)} & \textbf{Major Progress (Breakthrough)}  \\ \hline
Number of Cepheid hosts & 37 SNe Ia hosts  & --- & 24 (not limited to SNe Ia hosts) & $\sim$100 \\ \hline
Max distance of Cepheid hosts & $\sim$40-50 Mpc & --- & --- & 100 Mpc \\ \hline
Number of geometric anchors & 4 (MW, LMC, SMC, NGC 4258) & --- & 6 (current + M31, M33) & 10 (current + M31, M33, NGC 6822, IC1613, WLM, IC10) \\ \hline
Total uncertainty on geometric anchor distances  & 0.6\% & --- & 0.5\% & 0.3\% \\ \hline
Number of distance indicators observed per galaxy & 1 & 2 (Cepheids + J-AGB) & 2 (Cepheids + TRGB) & 3 (Cepheids, TRGB, \& J-AGB measured self-consistently for most/all galaxies) \\ \hline
\end{tabular}
    \caption{Physical parameters necessary to achieve a 1\% $H_0$ measurement with a two-step distance ladder.}
    \label{tab:parameters}
\end{table*}

\section{Description of Observations}
\label{sec:observations}

\begin{figure*}[htb!]
    \centering
    \includegraphics[width=\linewidth]{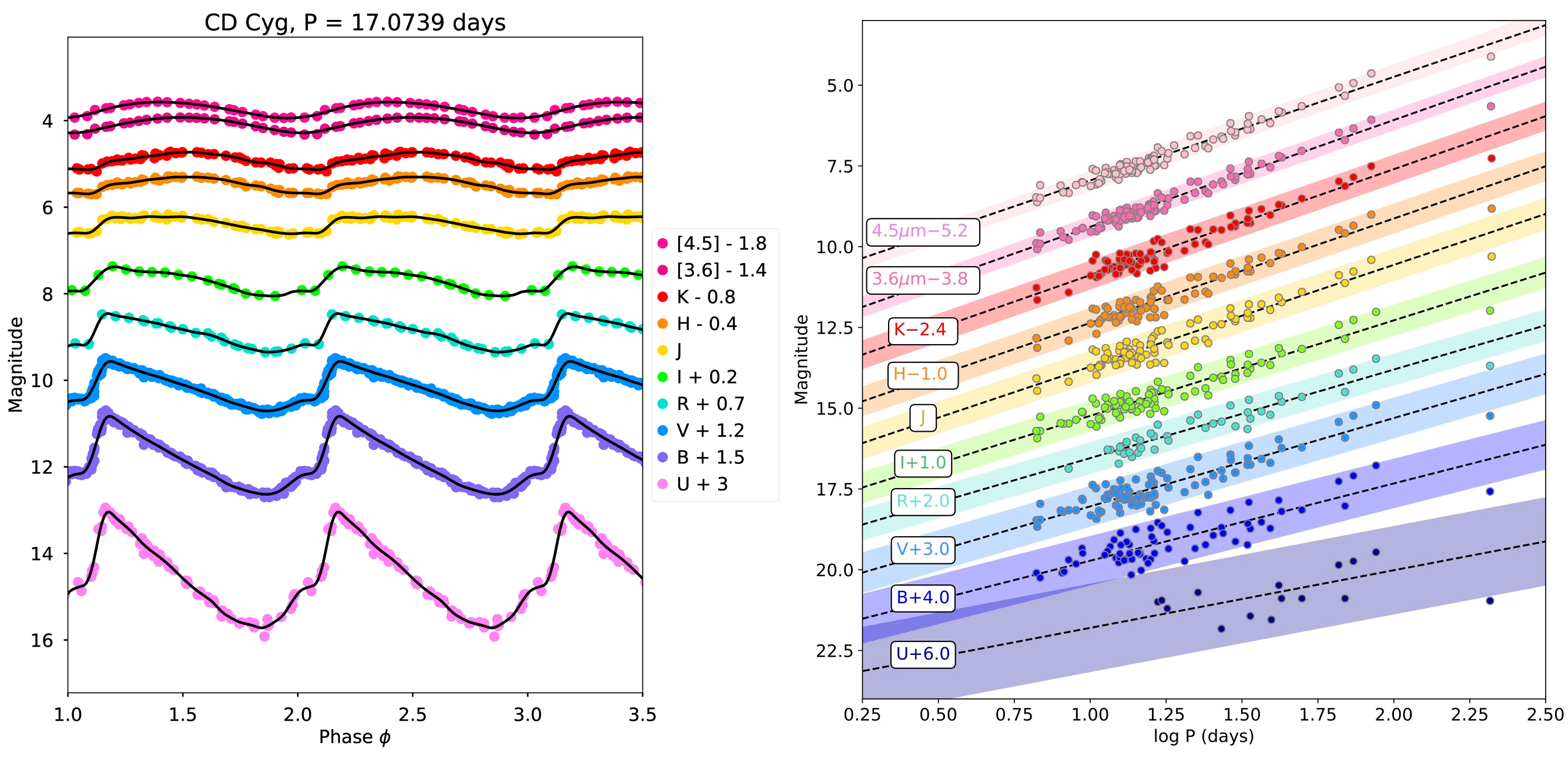}
    \caption{Example Cepheid light curves (left) and period-luminosity relations (right) in the blue-optical through near-infrared, reproduced from \citet{2016ApJ...816...49S} and \citet{scowcroft19}. The light curve amplitude is much larger at shorter wavelengths, which is optimal for detection and period determination, but the intrinsic PL scatter is lower in the infrared. RR Lyrae variables show similar light curve and PL characteristics.
    }
    \label{fig:multiband_light_curve_PL}
\end{figure*}

\subsection{Distance measurements in distant galaxies} \label{ssec:distant_distances}

A Cepheid with an average $M_V = -4$ (pulsation period $P > 10$ days) at 100 Mpc will reach $S/N=10$ in V with 2 hours of exposure time on a 6m HWO\footnote{\textbf{N.B.} We have made a small modification to the HWO exposure time calculator package, found at {\url{https://github.com/spacetelescope/syotools/}}. We follow the recommendations of \citet{2024ApJ...970...36S} for optimal photometric extraction aperture sizes for the JWST ETC, to match the S/N achievable with PSF photometry. The default extraction apertures of both ETCs, while suitable for bright and isolated sources, severely underestimate the S/N of faint sources relative to an optimal (PSF-weighted) extraction aperture. \citet{2024ApJ...970...36S} recommend an extraction aperture of no more than 0.75 times the PSF FWHM for JWST, as opposed to the 3 FWHM aperture implemented in \texttt{syotools}. Our reported exposure time calculations for HWO use this 0.75 FWHM extraction aperture, which is a much closer approximation of the photometry technique typically used for extragalactic resolved star imaging.}. Cepheid characterization requires 12 epochs over a 60-100 day interval (with nonredundant spacing) to adequately sample the full light curve for robust detection and period determination for a total of 24 hours per galaxy. The same amount of time will be needed in $I$ and/or $J$ to reach a precise period-luminosity relation, as there is lower scatter in the IR PL (see Figure 6), and color information is needed to use reddening-free Wesenheit magnitudes\footnote{Starting with \citet{1982ApJ...253..575M}, this magnitude system has been used to reduce the impact of dust in the host galaxy in the vicinity of the Cepheid.}.

For the TRGB at $M_I = -4$, one needs to reach a limiting magnitude $\sim$1.5 mag fainter than that in $I$ and one other band ($V$ is likely too faint, and $K$ is background-limited, leaving $J$ and $H$). At 100 Mpc this will take 25 hours in $I$ for a 6m aperture (about the same as the total Cepheid exposure time), but only about 10h in $J$ or $H$. $JH$ alone is not recommended, however, due to the narrowness of its color baseline and the steep NIR-TRGB color dependence \citep{2012ApJS..198....6D, 2020ApJ...898...57D, 2024ApJ...966..175N, 2024ApJ...975..195N}.

Only a 6m or greater HWO will have the angular resolution, sensitivity, and PSF stability to obtain precise photometry of individual stars in galaxies at 100 Mpc. \emph{Without these capabilities, a robust two-step Hubble Constant measurement will likely remain out of reach.}

\subsection{Geometric anchors} \label{ssec:geometric_anchors}

Ironically, the nearest geometric calibrators may pose the greatest difficulties for HWO due to their brightness; indeed, they are a challenge even for HST. Cepheids in the Magellanic Clouds ($\sim14 < m_V < 20$) are observable with subarrays on HST \citep{2024ApJ...973...30B}, which allow for shorter readout times than the full frame, and Galactic Cepheids ($7 < m_V < 10$) must be observed with spatial scans \citep{2018ApJ...861..126R}, which smear the flux over many more pixels than in the typical ``stare" observing mode. 
Saturation limits are dependent on detector characteristics (e.g. readout time, full-well capacity, linearity behavior) as much as on telescope aperture, so we do not yet have sufficient information to make concrete predictions for the treatment of bright sources.
However, we strongly recommend that workarounds for the nominal imaging saturation limit---such as neutral density filters, detector subarrays, scanning mode, grism disperser, and/or diffuser---be considered in the HWO design process. \emph{Without such modes, HWO will not be able to reach even the present systematic uncertainty floor for the nearest geometric anchors.}

It is crucial to note that establishing additional eclipsing binary anchors will likely require ELTs for spectroscopic follow-up, as current 8-10m class facilities can obtain radial velocities only for the brightest few EB systems at the distance of M31 \citep[$m_V < 20.5$;][]{2014ApJ...797...22L, 2019BAAS...51c.456B}. In the meantime, improvements in the surface brightness-color relation for early-type stars via Gaia parallaxes and new and upcoming interferometric studies may enable early-type EB stars to be measured with precision comparable to the late-type ones used in the LMC \citep{2017ApJ...837....7G, 2020A&A...640A...2S, 2021A&A...652A..26S}. 

\begin{figure*}[htb!]
    \centering
    \includegraphics[width=\linewidth]{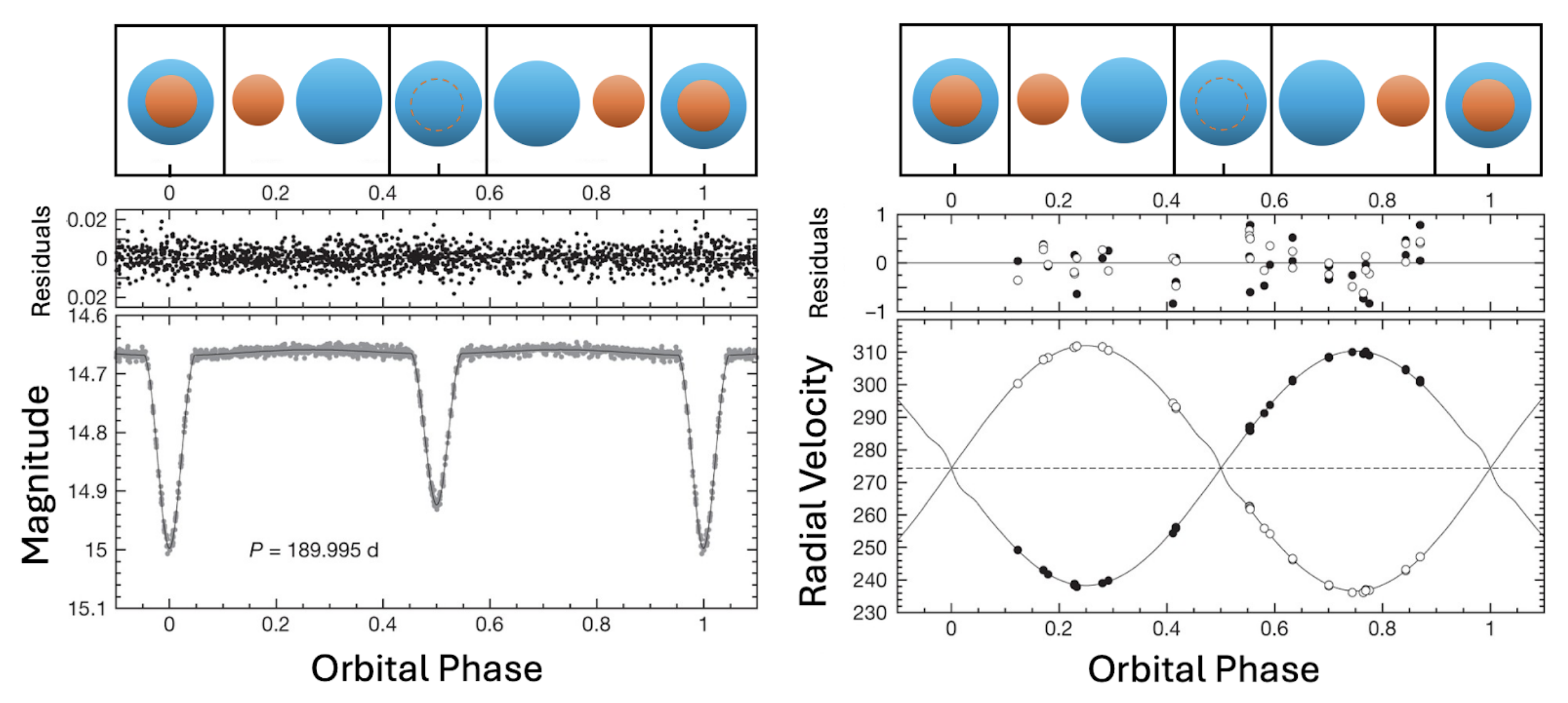}
    \caption{Geometric distances can be determined using spectrophotometric characterization of detached eclipsing binaries \citep[adapted from][]{2013Natur.495...76P}. The left panel shows a phased lightcurve with two eclipses, the right panel shows a phase velocity curve, and the cartoons show the state of the system through the phases (n.b., the relative sizes are exaggerated for illustrative purposes). These data derive the physical radius of each star, which is converted to a physical distance with a radius-surface brightness relation. The limitation for this work is in the spectroscopy, which must be of sufficient spectral resolution (and stability) to separate the two components to derive relative radial velocities. }
    \label{fig:eclipsing_binaries}
\end{figure*}

Alternatively, with the ultra-stable capabilities required to meet coronagraph pointing needs, it is also possible that HWO itself could provide spectroscopy of sufficient spectral resolution and pointing stability to characterize eclipsing binaries (see Figure~\ref{fig:eclipsing_binaries}). Such measurements would require optical spectroscopy at medium to high resolution that is comparable to other science drivers for the spectrograph.

\subsection{Additional considerations} \label{ssec:additional}

\begin{itemize}
    \item The inclusion of a long-pass ``white light" filter, such as F350LP on HST/WFC3-UVIS, would greatly increase variable detection efficiency in distant galaxies, but is not strictly necessary for these purposes.
    \item The ability to take simultaneous, co-spatial imaging in the optical $VI$ and infrared $JHK$ bands (analogous to JWST/NIRCam’s short- and long-wavelength channels) would be highly desirable, although again not strictly necessary. The introduction of a dichroic element would reduce the total telescope time needed for both Cepheids and the TRGB by a factor of $\sim$1.5-2 (assuming it would not otherwise impact overall transmission substantially). Furthermore, it would ensure that the J-AGB would be measurable at high precision with no additional exposures in effectively any Cepheid- or TRGB-optimized program (assuming sufficient statistics, which will depend on the imager’s FOV and the galaxy’s star formation history). SBF observations would benefit similarly from simultaneous multiwavelength coverage.
    \begin{itemize}
        \item On the FOV, a NIRCam-like FOV (2 $\times$ 5 arcmin) is sufficient to capture both Cepheids in the disk and the TRGB in the halo in a single pointing for most spirals outside the Local Group (as in e.g. NGC 4258 at 7 Mpc; JWST GO-1685, 1995, 2875). Efficiency gains from larger areas will have the most impact on the nearby anchors.
    \end{itemize}
    \item Like Cepheids, RR Lyrae have low amplitudes in the IR (only $\sim$0.1 mag in $H$), making JWST unsuitable for discovering them. They are faint enough ($M_V \approx +0.5$) that HST can detect them only within the Local Group (e.g., within 1-2 Mpc). A 6-meter HWO would be able to detect RRL out to 7 Mpc (distance modulus 29.2), reaching S/N=10 in $V$ in just over half an hour of exposure time. Although one could in principle observe them further out with longer exposure times, their short periods (0.3-1 day) necessitate high-cadence photometry for robust detection and period determination.
    \item Why not ELTs with adaptive optics?
    \begin{itemize}
        \item AO in the optical is much more difficult than in the NIR, especially bluer than the I band.
        \item FOVs of AO images are limited to an arcminute.
        \item PSF recovery has not been demonstrated to be stable enough for the precise and repeatable photometry required.
    \end{itemize}
\end{itemize}

\begin{table*}[htb!]
    \centering
    \begin{tabular}{|x|x|x|x|x|}
    \hline
\textbf{Observation Requirement} & \textbf{State of the Art} & \textbf{Incremental Progress (Enhancing)} & \textbf{Substantial Progress (Enabling)} & \textbf{Major Progress (Breakthrough)} \\ \hline
Imaging of resolved stars in distant galaxies & Highest available imaging resolution = JWST IR & JWST-resolution imaging in optical (0.031$\arcsec$ pixels vs. 0.04$\arcsec$ for HST) & 0.02—0.025$\arcsec$ pixels in optical (severely undersampled) & Well-sampled, diffraction-limited optical+IR imaging \\ \hline
Wavelength range & NIR & $VI$ & $VI$ & $VI$+NIR \\ \hline
Pixel scale & 0.031$\arcsec$ in NIR & 0.031$\arcsec$ in optical & 0.02-0.025$\arcsec$ in optical & 0.01$\arcsec$ in optical/NIR (Nyquist sampled in $V$) \\ \hline
Spatial resolution (PSF FWHM) & $\sim$0.035$\arcsec$ at 1 $\mu$m & $\sim$0.035$\arcsec$ in optical & $\sim$0.025$\arcsec$ in V & Diffraction limited at 500nm, $\sim$0.02$\arcsec$ (6m aperture) \\ \hline
FOV & 11 arcmin$^2$ (HST ACS), 10 arcmin$^2$ (JWST NIRCam) & 11 arcmin$^2$ across optical + NIR & 25 arcmin$^2$ across optical + NIR & 36 arcmin$^2$ across optical + NIR \\ \hline
Faint limiting magnitude in given band & F814W $<$ 29 (HST), F090W $<$ 31 (JWST) & $V < 31$, $I < 30$ & $V < 32$, $I < 31$ & $V < 33$, $I < 32$ \\ \hline
Bright limiting magnitude in given band & F090W $>$ 16 (JWST, stare mode), F555W$\sim$7 (HST, scan mode) & --- & $I > 12$ (Magellanic Cepheids) & $I > 7$ (MW Cepheids) \\ \hline
Absolute flux calibration uncertainty & 1\% (HST + CALSPEC) & --- & --- & 0.5\% (Landolt mission) \\ \hline
\end{tabular}
    \caption{Observational benchmarks for progress in measuring $H_0$.}
    \label{tab:observations}
\end{table*}

\section{Summary}
\label{sec:summary}

The anticipated capabilities of NASA's Habitable Worlds Observatory will allow us to extend measurements of galaxy distances via their resolved stars out to a much larger volume. Such measurements would allow for the construction of a two-rung distance ladder enabling a 1$\%$ measurement of the Hubble constant (without the use of Type Ia supernovae), as well as measurements of distances to low-mass dwarf galaxies via their RR Lyrae populations. Special care will need to be taken to ensure that HWO will be able to observe the very bright stars in the nearby geometric anchors.

%\acknowledgements
{\bf Acknowledgements.} 
RLB acknowledges support from NSF-AST 2108616. GSA acknowledges support from the STScI Director's Discretionary Fund.  MJD acknowledges support from HST AR-16611. The authors warmly thank the HWO Project for their support during this process.

\bibliography{author.bib}

\end{document}